\documentclass[onecolumn]{aa}
\usepackage{graphicx}
\begin{document}
   \title{Korean Nova Records in A.D. 1073 and A.D. 1074: R Aquarii}

   \author{Hong-Jin~Yang
          \inst{1},
          \ Myeong-Gu~Park
          \inst{2},
          \ Se-Hyung~Cho
          \inst{1},
          \and
          Changbom~Park \inst{3}
%          \fnmsep\thanks{Just to show the usage
%          of the elements in the author field}
          }

   \offprints{H.-J.~Yang, \email{hjyang@kasi.re.kr}}

   \institute{Korea Astronomy and Space Science Institute, 61-1 Hwaam, Yuseong, Daejeon 305-348, Korea\\
              \email{hjyang@kasi.re.kr, cho@kasi.re.kr}
         \and
             Department of Astronomy and Atmospheric Sciences,
             Kyungpook National University, Daegu 702-701, Korea\\
             \email{mgp@knu.ac.kr}
         \and
             School of Physics, Korea Institute for Advanced
             Study, Seoul 130-722, Korea\\
             \email{cbp@kias.re.kr}
             }

   \date{Received November 30, 2004; accepted December 31, 2004}

   \abstract{
   R Aqr is known to be a symbiotic binary system with an associated extended
   emission nebula, possibly produced by a historic outburst. To find the
   associated historic records, we searched for and compiled all ¡®Guest
   Star¡¯ and ¡®Peculiar Star¡¯ records in three Korean ¡®official¡¯
   history books that cover almost two thousand years, Samguksagi, Goryeosa,
   Joseonwangjosillok. In addition to the record of A.D. 1073, previously
   noted by Li (1985), we have found in Goryeosa another candidate
   record of A.D. 1074, which has the same positional description as that
   of A.D. 1073 with an additional brightness description. We examined
   various aspects of the two records and conclude that they both are
   likely to be the records of outburst of R Aqr. This means that there
   were two successive outbursts in A.D. 1073 and in A.D. 1074,
   separated by approximately one year.
   Based on these records, we estimate the distance to R Aqr to be 273 pc
   if the expansion of the nebula has been at a constant rate.
   The brightness record of A.D. 1074 corresponds to the absolute magnitude
   at outburst of $ M_{outburst} = -6.2^{m} \sim -5.2^{m}$
   at this distance. The two Korean records associated with R Aqr may
   provide astronomically meaningful constraints to the outburst model of R Aqr
   and the formative process of its nebulosity.
   }

   \titlerunning{Historical records of R Aquarii}
   \authorrunning{H.-J.~Yang et al.}
   \maketitle
%
%________________________________________________________________

\section{Introduction}

   R Aquarii is one of the more enigmatic astronomical objects
   that has attracted investigators for many decades. R Aqr,
   located at $RA\ 23^{h}\, 43.8^{m},\ Dec. -15\degr \, 17.0\arcmin$ (J2000.0),
   is the nearest known symbiotic binary system at about 250 pc
   distance (Whitelock 1987) and also has an astrophysical
   jet (Paresce et al. 1991). The primary is a Mira variable of
   spectral type M7e with a pulsational period of 386.83 days
   whose variable nature was first noted by Harding (1816).
   The secondary is generally believed to be a white dwarf
   (Henny \& Dyson 1992). R Aqr is known to be associated with
   an extended emission nebula (Herbig 1965). The complex nebula
   which surrounds the central star in a remarkable point-symmetric
   arrangement consists of two main components, the ¡®inner¡¯ and ¡®outer¡¯
   nebulosities, both discovered by Lampland (1923a,\,b).
   The outer nebulosity is an oval-shaped formation of about 90\arcsec \ extent
   composed of two intersecting, well-defined arcs and shape of a double
   convex lens oriented nearly east and west (Solf \& Ulrich 1985). Outward
   motions in the outer nebulosity were suspected by Hubble (1940, 1943).
   Baade (1944) detected outward motions of the nebula from comparison
   of plates taken 16 years apart and concluded that, assuming a constant
   rate of expansion, the nebula was 600 yr old. The inner nebulosity
   shows a wedge-shaped structure, that reaches about 30\arcsec \ extent and
   appears to extend north and south, perpendicular to the orientation
   of the outer nebulosity (Solf \& Ulrich 1985). Two nebulosities have
   dense equatorial ring structure. Equatorial velocities of 55 and 32
   $\mathrm{km\,s^{-1}}$ are derived from the outer and inner shell-like
   nebulosities, respectively, and the expansion rates combined with
   tangential motions from the literature yield a distance of about
   180 pc for both shells (Solf \& Ulrich 1985).
   Solf and Ulrich (1985) argued that the uniformity observed in the kinematics
   of each shell indicates that the shells can be attributed to two major
   ejection events, 650 and 190 years ago, and that further outbursts
   within the last centuries are unlikely.

   A new feature within the inner nebulosity appeared between 1970 and 1977.
   A spike or jet has appeared as a protrusion from the central star
   toward the northeast (Hollis, Oliversen \& Wagner 1989).
   The jet has been observed in the optical (Wallerstein \& Greenstein 1980; Herbig 1980),
   in the radio at 2 and 6 cm with the VLA (Sopka et al. 1982;
   Kafatos \& Michalitsianos 1983), and in the far UV with the
   International Ultraviolet Explorer (IUE). The jet may be powering the
   R Aqr nebula that radiates more than $5 \times 10^{44} \
   \mathrm{erg\,yr^{-1}}$
   in the Balmer and Lyman continua and emission lines (Kafatos \& Michalitsianos 1982).

   Kafatos and Michalitsianos (1982) and Li (1985) separately suggested
   historical records related to the outburst of R Aqr. Kafatos and
   Michalitsianos (1982) proposed that a historical Japanese astronomical
   record of a nova outburst in A.D. 930 might be associated with R Aqr,
   which formed the outer extended nebulosity. However, Li (1985) pointed
   out the erroneous identification of the Japanese record of A.D. 930 for R Aqr.
   The Japanese record was based on records in Meigetuki and Ichidaiyouki,
   reproduced later in Japanese Ancient Astronomical Records
   (Kanda 1935). The record in Ichidaiyouki reads:¡°A Guest Star
   enters Urim (one of the oriental constellations), begins to move
   westwards in Urim.¡± Li (1985) concluded that in view of the words ¡®enter¡¯
   and ¡®move westwards¡¯, the event of A.D. 930 should be considered as that
   of a comet rather than a nova. We confirm this because the record of 930
   is actually classified as a comet in Japanese Ancient Astronomical Records
   (Kanda 1935). Li also suggested a Korean astronomical record of A.D. 1073
   as a description of nova outburst of R Aqr. The Korean record of 1073 reads
   simply:¡°A Guest Star is seen south of DongByeok (another oriental
   constellation, located in the north of Urim).¡±

   Historical nova observations have been recorded in many countries such as Korea,
   China and Japan as well as various European and Arabian countries. Lundmark (1921)
   collected many observations of historical novae and compiled a catalog that includes
   60 suspected nova records before or during 19th century. Most of Lundmark's (1921)
   catalog is based on Chinese records translated by Biot (1846) and Williams (1871).
   Hsi (1958) revised the previous list of historical Chinese ¡®New Star¡¯ catalog of 90
   suspected novae. While Korea has abundant suspected historical nova records,
   they have not been compiled systematically from the original sources in the previous
   literatures. Hence, we searched for and compiled all ¡®Guest Star¡¯ and ¡®Peculiar
   Star¡¯ records in the original ¡®official¡¯ Korean chronicles.
   We have found two records that could be associated with the outburst of R Aqr:
   records of A.D. 1073 and 1074 (Cho et al. 1999). The record of A.D. 1074
   that even has brightness information has never been mentioned as a candidate
   record of outburst of R Aqr.

%__________________________________________________________________

\section{Korean Records of Guest Star and Peculiar Star}

   Korea has abundant and homogeneous historical astronomical records in the
   three major history books: Samguksagi\footnote{All Korean historical names
   in this paper follow the notation used by the Korean government. Corresponding names
   in Chinese characters are given in the Appendix A.}
   (The History of the Three Kingdoms), Goryeosa (The History of the Goryeo Dynasty),
   and Joseonwangjosillok (The Annals of the Joseon Dynasty). They are deemed to
   be ¡®official¡¯ history books because the subsequent dynasties
   or the reigning monarchy itself oversaw
   the writing based on previous history books. Samguksagi covers the period
   from 57 B.C. to A.D. 935: the period of Three Kingdoms, namely Silla,
   Goguryeo, and Baekje. It contains total of 236 records of solar eclipses,
   comets, meteors, planet motions and so forth. Goryeosa covers the Goryeo dynasty
   from A.D. 918 to A.D. 1392. It is a well-arranged history book divided into several
   chapters by contents, and records are listed in chronological order.
   Joseonwangjosillok (Sillok, hereafter), the most recent and the most extensive official
   chronicles, covers Joseon dynasty from A.D. 1392 to A.D. 1910.
   In addition to these official history books, Jeungbomunheonbigo,
   a collection of the rearranged records compiled from the official Korean chronicles,
   also has abundant astronomical records divided into several chapters by contents.

   Korean historical astronomical records are in general described as one of
   several classes of astronomical phenomena such as solar and lunar eclipses,
   comets, meteors, planet motions and so forth. Among these records, some of
   ¡®Guest Star¡¯ and ¡®Peculiar Star¡¯ records are possibly related
   with outburst of stars.
   Literally, Guest Star means a new star. The first Korean Guest Star record appears
   in A.D. 85 and the second in A.D. 154, both in Samguksagi. More Guest
   Star records appear continuously up to A.D. 1770. We have compiled all Guest Star
   records in Table 1 from the Samguksagi, Goryeosa, Sillok, and Jeungbomunheonbigo,
   covering almost two thousand years. We found 39 of them. However, two records of
   A.D. 1600 seem to be duplicate of the same record in 1604. The two records in 1600
   appear only in Jeungbomunheonbigo which is a encyclopedia, and sometimes contains
   inaccurate records, particularly in date. Furthermore, the descriptions of two
   records in 1600 are the same as that in 1604.

\begin{table}
\begin{minipage}[t]{17cm}
\caption{Historical Korean Guest Star records compiled from the
         three Korean history books (Samguksagi, Goryeosa, and
         Joseonwangjosillok) and an encyclopedia (Jeungbomunheonbigo).}
\label{table: 1}
\centering
\renewcommand{\footnoterule}{}  % to avoid a line before footnotes
\begin{tabular}{lclccc}
\hline\hline

Date of observation\footnote{Julian calendar for dates prior to 4
Oct. 1582, Gregorian calendar after 15 Oct. 1582.}
\footnote{Records with no month or day information are noted by
¡®--¡¯.}
              &  JD\footnote{The record with no day information is
set to be the fifteenth day of recorded lunar month.}
                       & Associated\footnote{Estimated position over the
 modern constellations deduced from original description on
 position described as oriental constellations, a planet or the
 moon.} & Ref.\footnote{S means Samguksagi, G Goryeosa,
 J Joseonwangjosillok, and M Jeungbomunheonbigo.}
& Duration\footnote{Observational duration for the same guest star}
& Note\footnote{Estimated astronomical phenomenon deduced from
original description} \\
 &  & constellations &  &  (days)  &  \\

\hline

-- May/June 85 & 1752255  &  Dra/UMi/Cep  & S,\,M &   &  \\ --
Jan. 154    & 1777321  &  --           & S,\,M &   &  \\ --
Oct./Nov. 299 & 1830571 & --           & S,\,M &   &  \\ -- -- 622
& --       &  --           & M     &   &  \\ -- Dec./Jan 867/868 &
2038094 & Lib/Sco/Sgr  & S,\,M &   &  \\ 01 Aug. 1065    & 2110262
&  --           & G,\,M &   &  \\ 10 Sep. 1073   & 2113224  &
Aqr/Psc/Cap   & G,\,M &   &  \\ 19 Aug. 1074   & 2113567  &
Aqr/Psc/Cap   & G,\,M &   &  \\ 10 Aug. 1163   & 2146065  & Oph
& G,\,M &   &  \\ 03 May  1356   & 2216460  & Tau           &
G,\,M &   &  \\ 09 June 1363   & 2219053  & --            & G,\,M
&   & Shower \\ 22 May 1391    & 2229262  & Dra/UMi/Cep   & G,\,M
&   &  \\ 05 Jan. 1399   & 2232047  & Oph           & J,\,M &   &
\\ 11 Mar. 1437   & 2245992  & Sco           & J,\,M & 14  &  \\
24 Aug. 1499   & 2268803  & Dra/UMi       & J,\,M & 4 & Comet \\
06 Nov. 1572   & 2295541  & Cas           & J,\,M &   &  \\ 23
Nov. 1592   & 2302853  & Cet           & J,\,M & 457  &  \\ 30
Nov. 1592   & 2302860  & Cas           & J,\,M & 118  &  \\ 04
Dec. 1592   & 2302863  & Cas           & J,\,M & 115  &  \\ 12
Dec. 1592   & 2302872  & And?           & J &   &  \\ 18 Jan. 1593
& 2302909  & And?           & J &   & ? \\ -- Nov./Dec. 1600 &
2305782 & Sco         & M &   & Typo\footnote{Typographical
error}? \\
-- Dec./Jan. 1600/1601 & 2305813 & Sco    & M &   &
Typo? \\ 13 Oct. 1604   & 2307195  & Sco           & J,\,M & 201
&  \\ 27 Oct. 1639   & 2319992  & Lep           & J     & 3  &
Comet \\ 04 Feb. 1661   & 2327763  & Del/Aql       & J     & 5  &
Comet \\ 13 Dec. 1661   & 2328075  & Aqr           & J,\,M & 20 &
Comet \\ 20 Nov. 1684   & 2336453  & Vir           & M     &    &
\\ 22 Apr. 1702 & 2342814  & Sco/Sgr       & J,\,M & 9  & Comet \\
13 Feb. 1737 & 2355530  & Peg/Aqr/Cep   & J     &    &  \\ 06 Mar.
1742 & 2357377  & Sgr/Cap       & J     & 2  & Comet \\ 13 Feb.
1743 & 2357721  & Crv-UMa-Crt   & J,\,M & 9  & Comet \\ 06 Jan.
1744 & 2358048  & Peg/And/Aqr   & J     & 2  & Comet? \\ 17 Nov.
1744 & 2358364  & Vir           & J     & 2  & Comet \\ 02 May.
1748 & 2359626  & Peg           & J     &    &  \\ 19 Dec. 1759
& 2363874  & Cet           & J     &    &  \\ 08 Jan. 1760   &
2363894  & Cma-Cet       & J,\,M & 14 & Comet  \\ 09 Feb. 1760   &
2363926  & Leo           & J,\,M & 11 & Comet \\ 29 June 1770   &
2367719  & Oph-Cap-Sgr   & J,\,M & 4  & Comet \\

\hline

\end{tabular}
\end{minipage}
\end{table}

   A Guest Star in Korean history books mainly represents a nova or a comet.
   However, when the movements of a Guest Star are subsequently observed,
   the Guest Star is recorded as a comet afterwards. For instance, one record
   in Goryeosa on 4 April 1066 is described as ¡°A star, as big as the Moon,
   appears to the northwest, and suddenly becomes a comet.¡±

   Sometimes Guest Star records have descriptions of position and size.
   Since the position of Guest Star is described based on the oriental constellation
   or relative to planets or moon, we can convert the position into the corresponding
   modern constellation. Ahn et al. (1996) and Park (1998) have identified most of oriental
   constellations with associated modern constellations. We have found five Guest Star records
   during the Three Kingdoms period. Two of them, records in A.D. 85 and A.D. 867/868,
   have positional information but the locations do not coincide with those of R Aqr.
   Goryeosa has seven Guest Star records. One in 1363 is recorded as a Guest Star,
   but it seems to be related with meteor shower. The description reads:¡°Seven Guest
   Stars appear simultaneously and three small stars fight against each other.¡±
   All records have positional information except the record of 1065. Among them,
   two records in 1073 and 1074 seem to be related with R Aqr based on estimated location and
   actual visibility on the recorded night.
   The record of 1073 has been suggested to be associated with the outburst of R Aqr by
   Li (1985). In addition to the record of 1073 we have found another new candidate record for
   the outburst of R Aqr. The positional information of the record of 1074 is the same
   as that of 1073, but it has an additional description of brightness.
   The description of records of 1073 and 1074 are shown in Table 3.

   Joseon dynasty has roughly 340 Guest Star records in total, describing about 25 independent
   Guest Star events. Most of them are the observations of the well-known Kepler's supernova
   of 1604. Twelve of them seem to be related with comets because the descriptions
   include the movement of the Guest Stars. These are marked as a ¡°Comet¡± in the seventh
   column in Table 1. However, we cannot be sure about the real nature of the object for
   every record. We also identified the location of all Guest Star events with associated
   constellations in Table 1. Among these records, only one record of January 1744 in Sillok has a
   possibility to be associated with R Aqr, it reads simply:¡°A Guest Star appears
   in the region of DongByeok and it's appearance resembles a comet.¡±
   The record of January 1744 also appears in Chinese and Japanese chronicles,
   but it is described as a comet in the two chronicles. Another record of 1737
   is also related with the constellation of Aquarius, but estimated location of the record
   is apart from that of R Aqr.

   Supernova is also recorded as a Guest Star in the Korean history books.
   Since Koreans believed that a special astronomical event hints the future,
   Guest Stars generally were recorded carefully. Although we do not find in Goryeosa
   the record of Crab supernova in 1054, while China and Japan have records of
   Crab supernova, it could have been caused by political disorder
   in the early Goryeo dynasty, because not a single astronomical record appears
   in Goryeosa for the whole year of 1054. Meanwhile, two supernova records of Tycho's
   and Kepler's are carefully recorded as Guest Stars in Sillok. Tycho's supernova
   was discovered by Koreans on 6 November 1572 and sighted two days later by Chinese. Kepler's
   supernova is also recorded from 13 October 1604 to 23 April 1605 in Sillok as a Guest
   Star with details such as size, angular distance from Polaris, color and so forth.

   In addition to Guest Star records, another class of records that can be related with
   nova or cataclysmic phenomenon are ¡®Peculiar Star¡¯ records. We also have compiled and listed
   all Peculiar Star records from Samguksagi, Goryeosa, Sillok, and Jeungbomunheonbigo.
   Some of them also have positional descriptions. We have listed the Peculiar Star records
   in Table 2. However, we have not found any adequate record to be related with R Aqr from
   the Peculiar Star records in Table 2.

\begin{table}
%\begin{minipage}[h]{\columnwidth}
\caption{ Historical Korean peculiar star records compiled from the
         three Korean representative history books (Samguksagi,
         Goryeosa, and Joseonwangjosillok) and Jeungbomunheonbigo.
         All notations in this table are the same as in Table 1.}
\label{table:3}
\centering
\begin{tabular}{lclcc}
\hline\hline

Date of        &  JD  & Associated & Ref. & Note \\
observation    &      & constellations &  &      \\

\hline
-- -- 673      &  --     &  --           & S     &    \\
-- -- 744      &  --     &  --           & S,\,M &    \\
03 Oct. 932    & 2061747 &  --           & G,\,M &    \\
31 May 1021    & 2094129 &  Com/Vir      & G,\,M &    \\
04 Oct. 1031   & 2097907 &  Cnc          & G,\,M &    \\
25 Oct. 1063   & 2109616 &  Oph          & G,\,M &    \\
14 Oct. 1072   & 2112893 &  Peg          & G,\,M &    \\
04 Aug. 1082   & 2116474 &  UMi          & G,\,M &    \\
11 Apr. 1176   & 2150693 &  --           & G,\,M &    \\
05 Sep. 1382   & 2226081 &  Leo          & G,\,M &    \\
10 Nov. 1577   & 2297371 &  --           & J  & Comet \\
-- -- 1625     & --      &  --           & M  &       \\
-- Aug./Sep. 1626 & 2315187 & --         & M  &       \\
-- Oct./Nov. 1627 & 2315612 & --         & M  &       \\
-- Feb./Mar. 1645 & 2321942 & Cnc        & M  &       \\
04 Dec. 1707   & 2344866 & Del/Vul       & J,\,M      \\

\hline

\end{tabular}
%\end{minipage}
\end{table}

%___________________________________________________________________________________________

\section{R Aqr and Guest Star of A.D. 1073 and 1074}

\subsection{Guest Star records of A.D. 1073 and 1074}

   Among all Korean historical Guest Star and Peculiar Star records listed in Table 1 and 2,
   we have found two most relevant records in Goryeosa (also in Jeungbomunheonbigo),
   those of A.D. 1073 and 1074, that can be related with outburst of R Aqr. Fig. 1
   shows the original text of the two Guest Star records of A.D. 1073 and 1074 in Goryeosa.
   While the two records are separated by one year interval, the locations
   are described in the same way:¡°A Guest Star is seen at the south of
   DongByeok.¡± DongByeok is one of the 28 oriental constellations (also known as ¡®28 lunar lodges¡¯),
   which consists of $\alpha$ And and $\gamma$ Peg
   (Ahn et al.~ 1996; Park~1998). Although the two records of 1073 and 1074 are
   described in the same way, the dates of two records, recorded in lunar
   calendar with 60 cyclical days, are different. Moreover, a sighting of
   Venus at daylight was also recorded along with the Guest Star
   record of 1073 for the same day. The apparent magnitude of Venus at
   10 September 1073 was $-4.^{m}19$ whereas the mean apparent magnitude of
   Venus is $-3.^{m}19$. These strongly support that the records of 1073 and
   1074 are real and independent.

\begin{table}
\caption{Guest Star records of A.D. 1073 and 1074}
\includegraphics{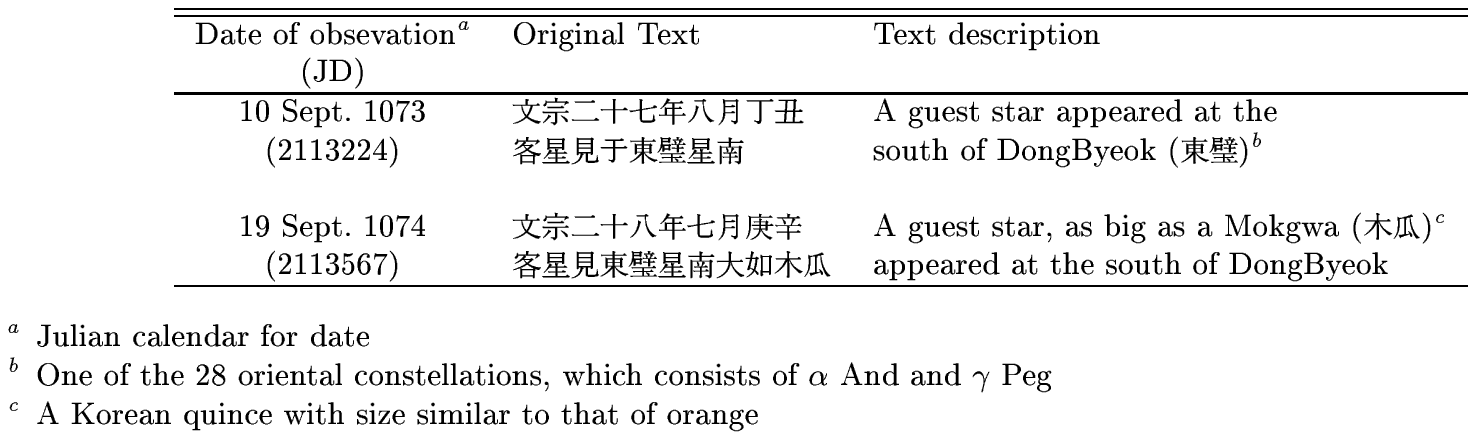}
\end{table}

   Also, notable is that the record of 1074 includes the size information of
   the Guest Star. The size of the Guest Star is recorded as a Mokgwa.
   Mokgwa is a Korean quince, the size of which is similar to that of the orange.
   Mokgwa (hereafter, quince) is often used to represent the size, i.e. brightness,
   of astronomical objects or phenomena in Korean history books. For instance,
   91 meteor and 2 comet records in Goryeosa are described having the size of quince.

   \begin{figure}
   \centering
      \includegraphics[width=8 cm]{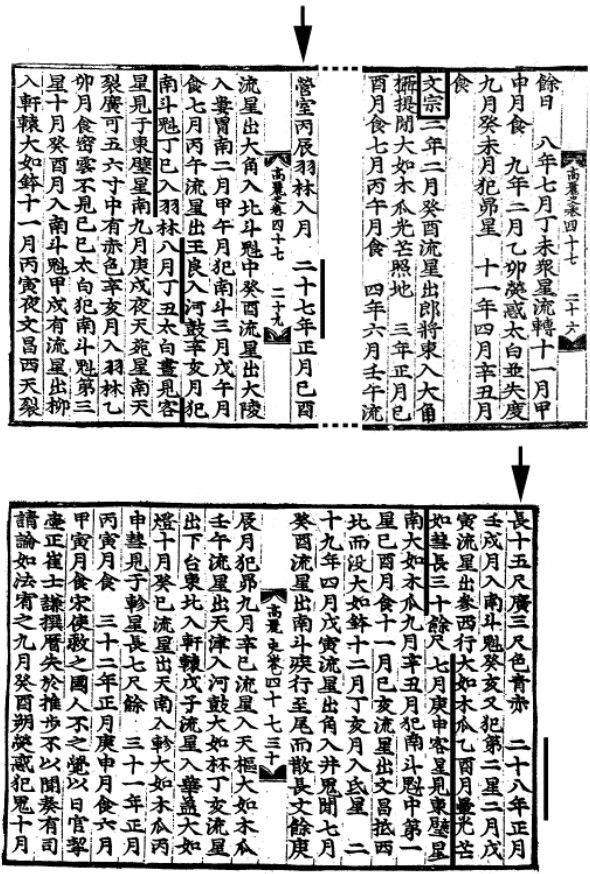}
     \caption{The original two records of guest star on September 10, 1073
            and August 19, 1074. The record of A.D. 1073 (in the upper box)
            reads as\,¡°A Guest Star is seen south of DongByeok,¡±
            and the record of A.D. 1074 (in the lower box) reads as
            ¡°A Guest Star is seen south of DongByeok and its size is
            similar to that of Mokgwa.¡± DongByeok is one of the
            ancient oriental constellations and corresponds to $\alpha$ And
            and $\gamma$ Peg. Mokgwa is a quince and its size is similar to that of orange.}
    \label{FigVibStab}
    \end{figure}

    The two records of Guest Star in 1073 and 1074 appear only in Korean history books.
    We cannot find similar records for the same period either in General Compilation
    of Chinese Ancient Astronomical Records (Beijing Observatory 1988)
    or in Japanese Ancient Astronomical records (Kanda 1935).

%________________________________________________________________________________

\subsection{Locations and Brightness}

   Li (1985) suggested the historical Korean record of A.D. 1073 as a record
   of nova outburst of R Aqr. Li, however, did not discuss the location of the
   Guest Star in 1073 relative to that of R Aqr.

   Oriental astronomers divided the visible part of the heavens into 31 sections,
   28 of which have been termed ¡®stellar divisions¡¯, and generally denoted by
   the asterism, forming the central or principal part of the division.
   Twenty-eight of them, each of which generally consists of several constellations, are
   located along the zodiac, whereas three of them, the ¡®Three Domains¡¯,
   occupy the central parts of celestial northern pole. The name of each division is
   also used for the representative constellation within it. We simply call these
   28 divisions as 28 oriental constellations, hereafter.
   The 28 oriental constellations are irregular in their extent, both from
   north to south and from east to west. Fig. 2 shows the location of R Aqr
   and constellation of DongByeok as appeared in Cheonsangyeolchabunyajido,
   the Korean sky map carved in stone in A.D. 1395, and in modern sky map.
   In terms of the oriental sky map, R Aqr is located in Urim, a constellation
   located in the south of DongByeok that is one of the 28 oriental constellations.
   Urim consists of 45 stars and occupies the region of
   $RA\ 20^{h}~ 30^{m} \sim  23^{h}\, 20^{m},\ Dec. -20\degr \sim 0\degr$.
   As already mentioned, the two Guest Stars are recorded as seen in the
   south of DongByeok. Since it is essential to compare the locations in
   two records with that of R Aqr, we made some pertinent investigations.

 \begin{figure*}
   \centering
     \includegraphics[width=14 cm]{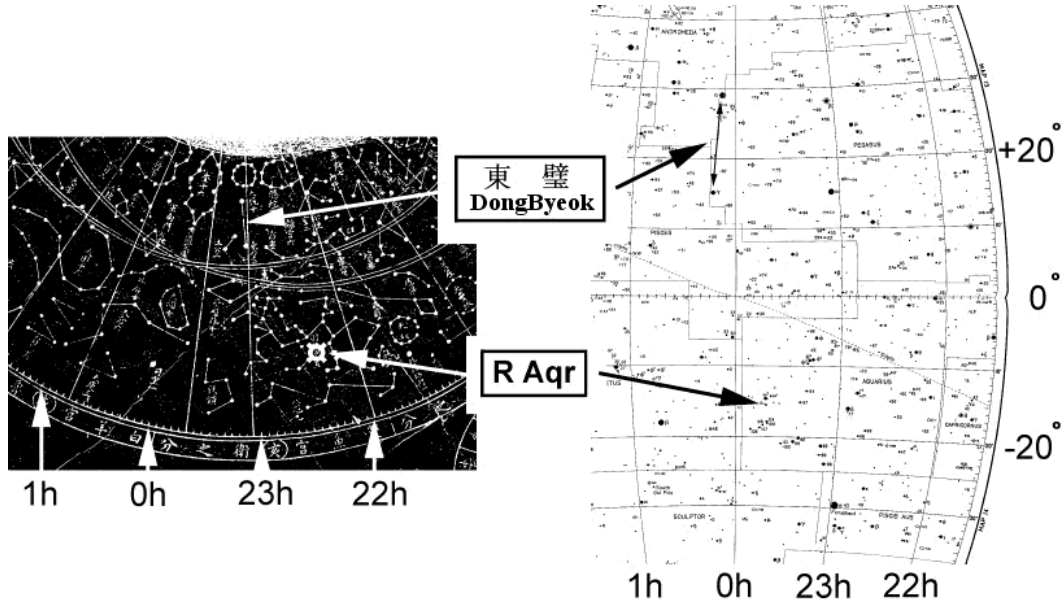}
   \caption{The position of R Aquarii and DongByeok. The sky map on the left, which is a part of
            Cheonsangyeolchabunyajido, a Korean sky map carved in stone
            in A.D. 1395, shows the position of DongByeok and a corresponding position of
            R Aqr. The sky map on the right shows a position of R Aqr and the position of
            corresponding DongByeok, $\alpha$\, And and $\gamma$\, Peg, in modern celestial coordinates.}
     \label{FigVibStab}
    \end{figure*}

   First, we checked the recent novae, supernovae, and cataclysmic variable
   stars within $40\degr \times 40\degr$ in the south of DongByeok that would cover the
   region from $\gamma$ Peg to the horizon at the time of $\gamma$ Peg's transit.
   We have not found any candidate nova or supernova. The frequency of the nova
   outbursts brighter than 2nd mag. recorded during the past one hundred years
   is estimated to be about 9 novae per century in Our Galaxy (Payne-Gaposchkin 1964).
   The probability of finding a random nova within $40\degr \times 40\degr$ during the past
   one thousand years is only about 0.32 (Kafatos \& Michalitsianos 1982).
   Hence, the probability that the outburst of R Aqr is not recorded and the Guest
   Stars of 1073 and 1074 are the records of yet another nova or supernova is even smaller.

   Second, we examined why the location of Guest Star is described as south of DongByeok
   instead of within Urim, the corresponding oriental constellation that encompasses R Aqr.
   We first searched recording formats of historical Guest Star records. Although we found
   no more Guest Star record with positional description in Goryeosa, we noted that the
   location of Guest Stars in Korean and Chinese chronicles are generally described with
   reference to the 28 oriental constellations such as DongByeok,
   which would have more significant astrological meanings than minor constellations such as Urim.
   For example, Guest Star records of A.D.\, 1604 $\sim$ A.D.\, 1605, Kepler's supernova,
   describe the location by the angular distance from the 28 oriental
   constellations and Polaris instead of the nearby minor constellations.
   And most Guest Star records in other official Korean history books have positional
   information based on the 28 oriental constellations. We also investigated the recording
   formats of the twelve Chinese Guest Star records from A.D. 1000 to A.D. 1300.
   Ten of them describe the location based on the 28 oriental constellations.
   In particular, the records of A.D. 1054, Crab supernova, appear independently in the
   four Chinese history books. Only one history book describes the location based on an
   ordinary star, whereas the positions in two history books are based on one of the 28
   oriental constellations, which is 30 degrees away from the Crab supernova. So we believe
   that the Guest Star records of 1073 and 1074 followed the astronomical tradition in Korean
   and Chinese chronicles that described their celestial positions based on the 28 oriental
   constellations where the event instead of nearby minor constellations.

   Moreover, it is quite likely that a brighter star was used as a reference
   when the location of a new or peculiar object was recorded. Magnitudes of stars in DongByeok
   ($\alpha$ \, And: $2.06^{m}$, \ $\gamma$\, Peg: $2.83^{m}$) are brighter than those in
   Urim ($3.27^{m} \sim 5.08^{m}$). Furthermore, when R Aqr passes the transit, Urim would
   be located at about $+27$ degrees above the horizon while DongByeok is located at about
   $+57$ degrees. If we consider the effect of extinction by air mass, the visible magnitudes
   of the brightest star in DongByeok and Urim are $2.24^{m}$ and $3.54^{m}$, respectively.
   In terms of brightness, the DongByeok might be more adequate objects than Urim as the
   reference star.

   Considering all aforementioned points, although the locations of the Guest Stars in 1073
   and 1074 are not described by the nearest constellation to R Aqr, we conclude that the identification
   of the Guest Stars of 1073 and 1074 with the outbursts of R Aqr is rather secure.

 %______________________________________________________________________________________

\section{Implication for R Aqr}

   Korean Guest Star records of R Aqr in 1073 and 1074 may provide some helpful hints to
   understanding of enigmatic R Aqr. First, we estimate two main physical parameters:
   distance to R Aqr and the maximum brightness at the outburst of R Aqr.
   We then explore the implications on the proposed outburst models of R Aqr.

 \subsection{Distance and Brightness}

   Several studies have provided the distance to R Aqr. Baade (1943, 1944)
   deduced an expansion age of about 600 years based on a constant expansion rate
   of outer nebulosity, and suggested a kinematical distance of 260 pc. The assumption
   of the constant expansion rate was based on several observational plates of R Aqr for
   16 years. L\'{e}pine et al. (1978) on the other hand estimated 181 pc, assuming an absolute magnitude for
   the Mira of $-8.1^{m}$ at 4 $\mu$m observation. Solf and Ulrich (1985) calculated the
   kinematical distance of 180 pc by observing the equatorial expansion velocity
   of $55\ \mathrm{km\, s^{-1}}$ and the expansion age of about 640 years. Whitelock (1987) quoted a
   distance of 250 pc to the R Aqr from GCVS (General Catalogue of Variable Stars).
   Recent Hipparcos measurement of the distance to the R Aqr is to be in the range of
   $122 \sim 521$ pc from the trigonometric parallax of $5.07 \pm 3.15$
   milliarcsec (Hipparcos Catalogue 1997).

   If we adopt the epoch of outburst to be A.D. 1073, the earlier of the two records,
   and further assume that the outer nebulosity is expanding at a constant rate (Baade 1943)
   with the expansion velocity of $55\ \mathrm{km\,s^{-1}}$ for the equatorial outer nebulosity
   (Solf \& Ulrich 1985), the radius of outer equatorial nebulosity is estimated to be
   10826 AU. The angular equatorial radius of 42\arcsec (Solf \& Ulrich 1985) corresponds to the
   distance of R Aqr of 273 pc. This distance falls within the Hipparcos distance range
   and consistent with the value of Baade and GCVS. If the nebula is decelerating, 273 pc
   is the lower estimate.

   Meanwhile, Henney and Dyson (1992) suggested a Sedov-type equation of motion for
   the outer shell-like nebulosity of R Aqr:
    \begin{equation}
      R_S{(t)} = R_o (t/t_o)^{3/5} ,
    \end{equation}

   where $R_s$ is the radius of the outer shell, $R_o$ the present radius and
   $t_o$ the present time since the outburst.
   When we adopt the Korean record of 1073 and the current expansion velocity of
   $55\ \mathrm{km\, s^{-1}}$, Eq. 1 yields a distance of 450 pc, marginally
   consistent with Hipparcos error range, but a little bit too large a distance
   compared to the previous estimates. This may suggest that the expansion of
   the outer nebulosity has not been significantly decelerated.

   Now, we estimate the brightness of the outburst. The record of 1074 describes
   the ¡®size¡¯, i.e. brightness, of the Guest Star as a quince. In order to estimate
   the brightness of an object described as a quince, we searched in Goryeosa for all astronomical
   records that had been described to have the size of quince in Goryeosa. We found 94 records: 91
   are meteor records, 2 comet records, and one nova record. We estimated the brightness
   of a quince from the meteor records.

   In Goryeosa, there are total of 735 meteor records, and 236 of them have the ¡®size¡¯
   information. Meteor is the second most abundant phenomena recorded
   in Korean history books, and the brightness is described by the\,¡®size¡¯\,of
   various objects (Yang, Park \& Park 2005). The ¡®size¡¯ of meteors in Goryeosa
   is described by 8 objects: They are egg, cup, quince, bowl, basin, doe, jar,
   chopping board, in increasing size, and the rest. Each of them appears
   in Goryeosa as many as 25, 43, 91, 2, 25, 2, 37, 3, and 8 times, respectively.
   The remaining 8 records are described in different forms compared to others,
   and we do not consider them. Among meteor records, a quince is the third or
   fourth smallest object and the most frequent one. Meteor event generally can
   be seen on the ground when it is brighter than 5th magnitude. We figure that
   the record with no description is to be 4th or 5th magnitude, the egg 3rd or
   4th magnitude, the cup 2nd or 3rd magnitude, and so forth in consecutive order.
   Hence, we estimate that the size of quince corresponds to the apparent
   magnitude of $1^{m} \sim 2^{m}$. If the record of 1074 indeed describes the
   outburst of R Aqr with apparent magnitude of $1^{m} \sim 2^{m}$,
   the absolute magnitude of the outburst is estimated to be
   $M_{outburst} = -6.2^{m} \sim -5.2^{m}$ for the distance of 273 pc.

%____________________________________________________________________________________________________

\subsection{Implications for the outburst and nebulosity of R Aqr}

   Now we discuss implications for the formative processes of the inner and
   outer nebulosity. The essential information about R Aquarii is one outburst
   record in 1073 and another in 1074, i.e., two outbursts within one year.
   Moreover, the record of 1074 has additional ¡®size¡¯ information which the record
   of 1073 does not have. Thus the outburst of 1074 was likely to be
   brighter than that of 1073. In addition, no outburst since then has been recorded.

   Kafatos and Michalitsianos (1982) proposed a jet plus accretion disk system
   produced by the supercritical accretion from the Mira to the hot companion during
   the periastron passage of highly elliptical orbit to explain the jet observed.
   They also suggested a nova-like outburst 500 to 1000 years ago to form the extended nebula.
   Korean records confirm this outburst. However, two successive outbursts within one year
   interval indicate that the outburst of R Aqr may have been of different nature compared
   to the usual slow or recurrent nova outburst. The important question is now: Can there
   be an outburst mechanism that produces two successive outbursts with later one being brighter?

   Other interesting questions that may be raised are: How are the two outbursts related
   with the inner and outer nebulosity? It was probably one of the two outbursts or two
   bursts together that produced the outer nebulosity, which was the original cause to
   look for the outburst record. But can we rule out the possibility that each burst
   successively produced each nebula, the inner and outer one? The expansion timescale of
   the outer nebula is 500 to 1000 years, as previously mentioned. Approximate timescale
   of the inner nebula, estimated by dividing the angular size of $\sim 13\arcsec$ times the
   distance of 273 pc with the expansion velocity of $32\ \mathrm{km\,s^{-1}}$, is about 500 years
   (Solf \& Ulrich 1985). But the uncertainty of the estimate is quite large, and the
   timescale as large as one thousand years may not be ruled out. Hence we may imagine
   that the preburst of 1073 created smaller and slower inner ring-like nebulosity while
   the main brighter outburst of 1074 produced the larger and faster outer nebulosity in a different direction.

%___________________________________________________________________________________________

\section{Summary and Discussion}

   R Aquarii is known to be a symbiotic binary system and associated with an extended
   emission nebula with inner and outer shell-like nebulosities. Judging from the angular
   size of the nebula, the estimated distance, and the expansion velocity, R Aquarii was
   suspected to have had an outburst that created current nebulosity within past one thousand years.

   Although the Japanese Guest Star record in A.D. 930 was proposed as a candidate outburst
   (Kafatos \& Michlitsianos 1982), we confirm Li's suspicion that it is a comet record
   (Li 1985; Kanda 1935). Bearing in mind Li's suggestion that a Korean record of A.D. 1073
   be the actual record of the outburst of R Aquarii, we searched for and compiled all
   Guest Star and Peculiar Star records in the three Korean official history books
   that cover almost two thousand years. We have found two
   records of Guest Star that may be relevant to R Aquarii. One is the Guest Star record
   of A.D. 1073 that has been mentioned by Li (1985). The record reads ¡°A Guest Star appears
   at the south of DongByeok,¡± where DongByeok is the one of 28 oriental
   constellations. We also have found another Guest Star record in A.D. 1074 that has the
   exactly the same description, but has an additional brightness description. Both records
   do not appear in Chinese nor in Japanese records. Moreover, any adequate record to be related
   with R Aqr does not appear since then in the Korean history books.

   We first examined the positional information in the records. Both records of 1073 and 1074
   describe the position as the south of DongByeok. We searched for other possible nova or
   supernova remnant within $40\degr \times 40\degr$ south of DongByeok. We found none.
   The probability of having a random nova in the same region for the past 1\,000 years is
   roughly 1/3, which is consistent with the records.

   Although the actual position of R Aquarii is within other oriental constellation
   Urim (Y\"ulin in Chinese) rather than DongByeok, we find that it is natural in
   Korea and China to describe important astronomical phenomena like novae or supernovae
   with respect to 28 oriental constellations. In addition, stars in
   DongByeok were more than 1 magnitude brighter than those in Urim.

   By fixing the time of outburst to be A.D. 1073 or 1074, combined with the current
   nebular size and the expansion velocity, we estimate the distance to R Aquarii to
   be 273 pc under the assumption of constant expansion. General consistency of this
   distance with previous estimates suggests that the deceleration of the nebula has
   not been significant.

   The record of 1074 further reads:¡°¡¦ its size being as large as a quince.¡±
   By comparing the size of Korean quince against other objects used as a degree of
   brightness in numerous meteor records, we estimate the apparent magnitude of R Aquarii
   at 1074 outburst to be 1st to 2nd magnitude. At the distance of 273 pc, this gives
   the absolute magnitude of $-6.2$ to $-5.2$ magnitude for the outburst.

   Analysis of the two records in 1073 and 1074 indicate they are independent and
   not an erroneous duplicate. This leaves us interesting puzzles. How can there
   be two nova outbursts separated roughly by one year? It is also possible to interpret
   existence of brightness information for 1074 outburst while none for 1073
   outburst as the former being brighter than the latter. If so, is there an outburst
   mechanism that produces two successive outbursts with the later one being brighter?
   How are two outbursts related to two nebulosities, the inner and outer one?

   Outburst records of R Aquarii in Korean history books provide the key confirmation
   on our general understanding of the nature of R Aquarii and the formation history
   of its nebula. At the same time, it presents us interesting questions that are needed to be
   solved for better understanding of R Aquarii.

%____________________________________________________________________________

\begin{acknowledgements}
      This work is partly supported by Korea Science \& Engineering Foundation
      through Astrophysical Research Center for the Structure and Evolution of the
      Cosmos. S.-H. Cho was partially supported by the National
      Strategic Program of the Ministry of Science and Technology,
      Korea.
\end{acknowledgements}

\clearpage
\appendix
\section{Words in Chinese characters and their English expression}

\begin{figure}[h] \centering
\includegraphics{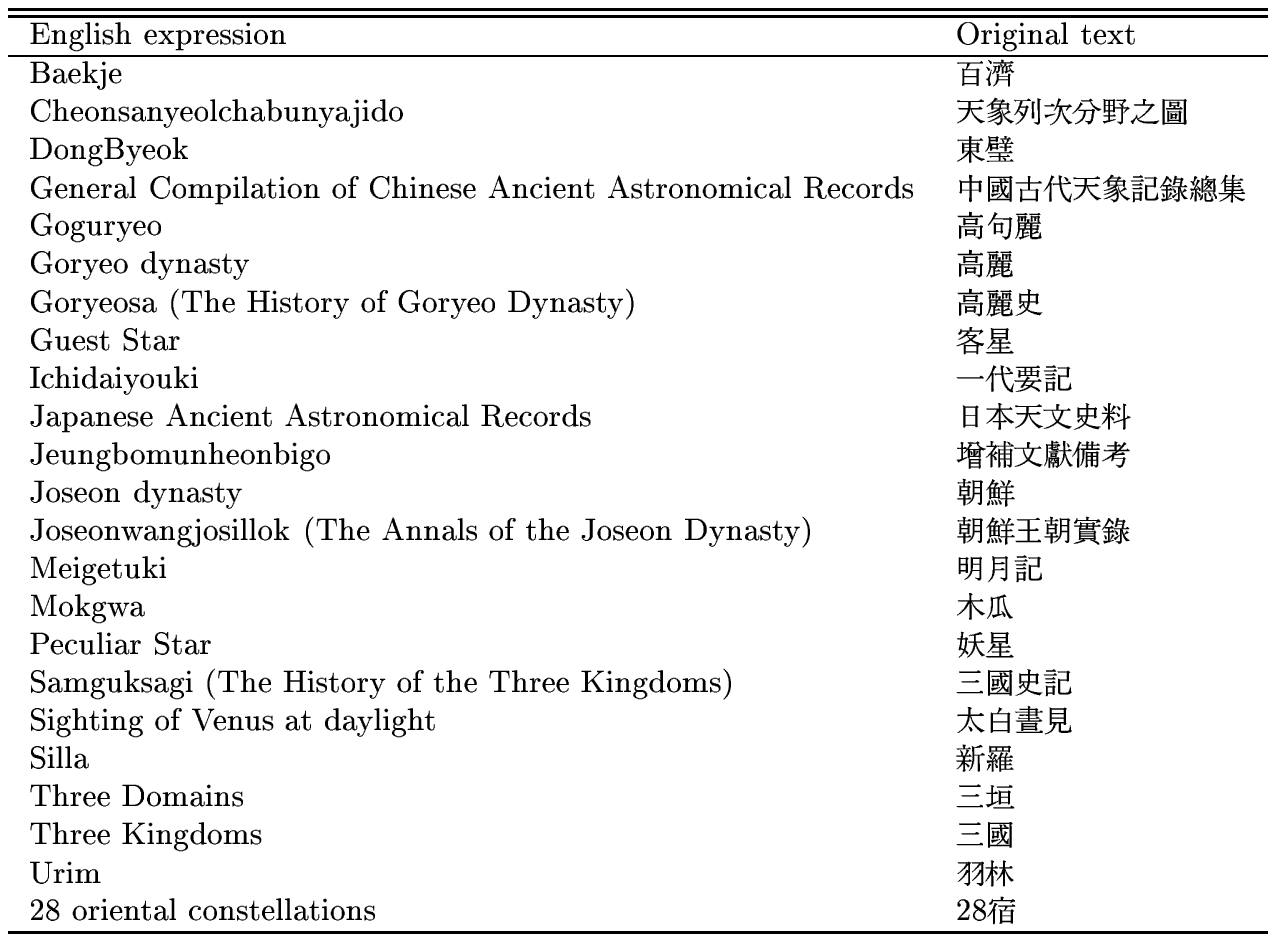}
\end{figure}

\end{document}